# Dynamical model for piezotronic and piezo-phototronic devices under low and high frequency external compressive stresses

Leisheng Jin[1], Xiaohong Yan[1], Xiangfu Wang[1] and Lijie Li[2*]

[1]School of Electronic Science and Engineering, Nanjing University of Posts and Telecommunications, China

[2]Multidisciplianary Nanotechnology Centre, College of Engineering, Swansea University, UK

**Abstract**
Dynamical theories for piezotronic and piezo-phototronic devices are incomplete. In this work, we aim to establish a theoretical method for modelling dynamic characteristics of devices exhibiting these two emerging phenomena. By taking the simplest piezotronic device—PN junction as an example, we develop a small signal model and the united approach to analyze its diffusion capacitance and conductance under both low and high frequency external compressive stresses, which is different from the traditional considerations that treat the piezopotential as a static value. Furthermore, we expand the theory into piezo-phototronic devices e.g. a light emitting diode (LED). The dynamic recombination rate and light emitting intensity are quantitatively calculated under different frequencies of external compressive stresses. The work complements existing works that only consider the static cases. The work can shed light in future high frequency piezoelectronic devices exploration.

**Key Words**：*diffusion capacitance, diffusion conductance, recombination rate, light emitting intensity*

## 1. Introduction

Piezotronics and piezo-phototronics, in recent years, have been witnessed rapid development[1]. From the invention of single ZnO (zinc oxide) nanowire generator [2] to the demonstration of piezoelectricity in single-layer $MoS_2$ [3]; from the first reveal of the coupling between piezoelectricity and photoexcitation in ZnO nanowires [4] to the successful application of using piezopotential in making active optoelectronics, such as luminescence devices for adaptive sensing [5], LEDs array for pressure imaging [6], photodetectors [7], solar cells [8] and so on, these two phenomena have been evolving toward more attracting research fields, bringing huge potential for the development of wearable electronics, robotics, the Internet of Things, biomedical engineering and human–machine interfacing [9, 10] [11]. These achievements, however, could not have been reached without understanding the fundamental physics behind. So far on the theoretical side, the frame work of piezotronics and piezo-phototronics is based on the semi-analytical abrupt junction model [12]. The first principle simulations of piezotronic transistors has been conducted, which can provide understanding for the width of piezoelectrically charged area [13]. In our previous work [14], a quantum scattering model was employed for investigating the ballistic carriers transportation in piezotronics. Meanwhile, numerical analyses by using tools such as DFT (density function theory) and FEM (finite element method) have also conducted to investigate the device interface physics and carrier's modulation on piezotronics [15] [16]. These developed theories have satisfactorily helped explain the existing experimental works [17] [18].

However, these theories or developed models focus mainly on the static study, where they tend to treat the piezopotential as a static value corresponding to a static compressive or tensile force. In terms of dynamic investigation, models for piezotronics and piezo-phototronics have not been established. The dynamic analysis for piezotronic and piezo-phototronic devices under a dynamic external force at a high frequency is usually essential in current research of piezotronic and piezo-phototronic devices, because these phenomena have been widely seen in experiments. For example, in work described in [6] and [17] , the authors fabricated an array of LEDs that can detect dynamic pressure and emit light accordingly based on piezo-phototronic effect, where the impact of the frequency of the exerted pressure is important, which is an unanswered problem. Moreover, the quantitative dynamic effects caused by the piezopotential on carrier's generation, recombination and transportation are not clear [19], although it has been always used by authors in explaining their experimental works [8, 20].

In this work, we develop a novel dynamic model for piezotronics and piezo-phototronics to investigate quantitatively how the dynamic force affects the carrier's generation and recombination process to explore the dynamic light-emitting mechanism in piezo-phototronic LEDs. We use the most basic piezotronic device – a PN junction in the analysis, trying to construct a general dynamic model for peizotronic devices subject to dynamic external forces. External forces at both low and high frequencies are considered using a small-signal model and unified approach [21]. Furthermore, we calculate the dynamic recombination rate and light emitting rate in piezotronic and piezo-phototronic LEDs, providing a thorough understanding of the dynamic process of carriers under the piezopotential. The work complements existing theories for piezotronics and piezo-phototronics [12] [10].

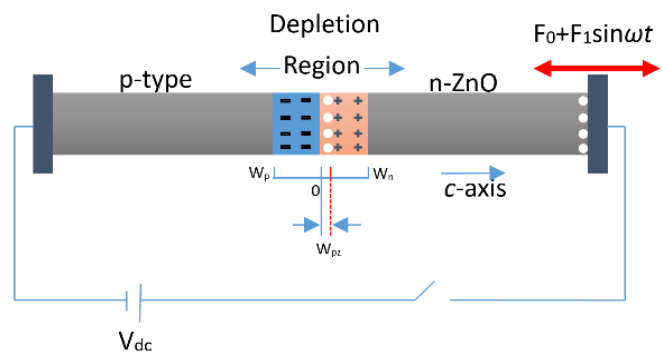

Figure 1. Basic structure of piezotronics: a typical PN junction with N part made by ZnO

**2. Basic theory**

The structure studied in this work, as shown in Figure 1, is a typical PN junction that is with n-part made by ZnO. Due to the piezoelectricity of ZnO, piezoelectric charges will be generated at the two ends of the n-part of the PN junction when a compressed or tensile force applied to the ZnO. The distribution of the piezoelectric charges is denoted by $W_{pz}$, taking $x=0$ at the interface of P and N region. The PN junction is assumed to be metallurgical. The depletion widths on the N and P regions are represented by $W_p$ and $W_n$, respectively. $V_{dc}$ is the forward applied bias. First, we consider the case when $V_{dc}$ is off and only static compressive force applied to the one end of the N region. The electrical filed across the depletion region of the PN junction can be easily derived, which are given by following equations (1-3).

$$E(x) = -\frac{eN_A(x+W_p)}{\varepsilon_s}, -W_p \leq x \leq 0 \tag{1}$$

$$E(x) = -\frac{e[N_D(W_n-x)+\rho_{pz}(W_{pz}-x)]}{\varepsilon_s}, 0 \leq x \leq W_{pz} \tag{2}$$

$$E(x) = -\frac{eN_D(W_n-x)}{\varepsilon_s}, W_p \leq x \leq W_n \tag{3}$$

where $e$ is elementary charge, $N_A$ and $N_D$ are donor and acceptor concentration, respectively. $\varepsilon_s$ is the permittivity of the ZnO. $\rho_{pz}$ is the density of polarization charges. As the electrical field $E$ is continuous at the metallurgical junction ($x=0$), by setting $x=0$ in equations (1) and (2), we have

$$N_AW_p = N_DW_n + \rho_{pz}W_{pz} \tag{4}$$

By integrating equations (1), (2) and (3), the potential φ(x) in depletion regions [-W_p 0], [0 W_p] and [W_p W_n] of the PN junction can be derived [12], and it is easy to know the built-in potential, as:

$$\varphi_{bi} = \frac{e}{2\varepsilon_s}(N_AW_p^2 + \rho_{pz}W_{pz}^2 + N_DW_n^2) \tag{5}$$

Combining equations (4) with (5), the built-in potential can be rewritten as:

$$\varphi_{bi} = \frac{e}{2\varepsilon_s}(N_AW_p^2 + \frac{N_A^2W_p^2}{N_D} + \rho_{pz}W_{pz}^2 - \frac{2N_AW_p\rho_{pz}W_{pz}}{N_D} + \frac{\rho_{pz}^2W_{pz}^2}{N_D}) \tag{6}$$

### 3. When time-variant compressive stress applied.

Now we assume there is a compressive stress varying as a sinusoidal function applied to the end of ZnO, i.e. $F=F_0+F_1\sin(\omega t)$, where $F_0$ are $F_1$ are amplitudes. $\omega$ is the frequency of the varying stress. $W_p$, $\rho_{pz}$, $N_A$, $N_D$ are all considered to be constants during one period of the stress added. Only $W_{pz}$ is varying with the same frequency as $F$, the static built-in potential is then replaced by:

$$\varphi_{bi}(t) = \frac{e}{2\varepsilon_s}\left(\left(N_A + \frac{N_A^2}{N_D}\right)W_p^2 + \left(\rho_{pz} + \frac{\rho_{pz}^2}{N_D}\right)(W_{pz0}^2 - 2W_{pz0}W_{pz1}\sin\omega t + W_{pz1}^2\sin^2\omega t) - \frac{2N_AW_p\rho_{pz}}{N_D}(W_{pz0} - W_{pz1}\sin\omega t)\right) \tag{7}$$

where $W_{pz0}$ is the width of piezoelectric charge region when only $F_0$ added, and $W_{pz1}$ is the width of piezoelectric charge region corresponding to $F_1$. The equation (7) can be rewritten as:

$$\varphi_{bi}(t) = c_1 + c_2 - c_3\sin\omega t + c_4\sin^2\omega t \tag{8}$$

by taking:

$$c_1 = \frac{e}{2\varepsilon_s}N_AW_p^2 + \frac{e}{2\varepsilon_s} \cdot \frac{N_A^2W_p^2}{N_D}; \quad c_2 = \frac{e}{2\varepsilon_s}\left(\rho_{pz} + \frac{\rho_{pz}^2}{N_D}\right)W_{pz0}^2 - \frac{e}{2\varepsilon_s} \cdot \frac{2N_AW_p\rho_{pz}}{N_D}W_{pz0}$$

$$c_3 = \frac{e}{2\varepsilon_s} \cdot \left(\rho_{pz} + \frac{\rho_{pz}^2}{N_D}\right) \cdot 2W_{pz0}W_{pz1} - \frac{e}{2\varepsilon_s} \cdot \frac{2N_A W_p \rho_{pz}}{N_D} \cdot W_{pz1}; \quad c_4 = \frac{e}{2\varepsilon_s}\left(\rho_{pz} + \frac{\rho_{pz}^2}{N_D}\right) W_{pz1}^2 \quad (9)$$

To have a clear picture of the dynamic built-in potential, we have calculated the $\varphi_{bi}$ in Figure 2. The parameters are taken as: $N_A$=1x10$^{16}$ cm$^{-3}$, $N_D$=1x10$^{15}$ cm$^{-3}$, $\varepsilon_s$=8.91, $W_{pz0}$=2x10$^{-7}$ cm and $W_{pz1}$=2x10$^{-8}$ cm. In determining $\rho_{pz}$ we have used the relation: $e_{33}s_{33}=e\rho_{pz}W_{pz0}$, in which $e_{33}$=1.22 C m$^{-2}$ and $s_{33}$=0.05%. The frequency of dynamical compressive force is taken as 1 KHz. From Figure 2, it is seen that the built-in potential varies with time in an approximate sine pattern. Moreover, we have shown the energy band changing under different compressive strains in Figure 3. The stains are changing from 0.002 to 0.1, it is seen the piezoelectric potential and the built-in potential $\phi_{bi}$ are both increased accordingly. It should be noted that due to the change of piezoelectric potential, the width of depletion region is also changed, contributed mainly from N-type region.

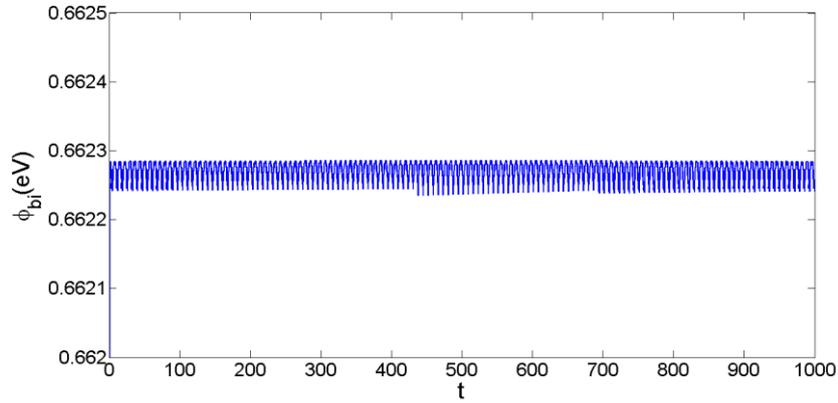

Figure 2. Calculation of dynamical built-in potential when there is a dynamical compressed force applied with frequency $\omega_p$=1 KHz.

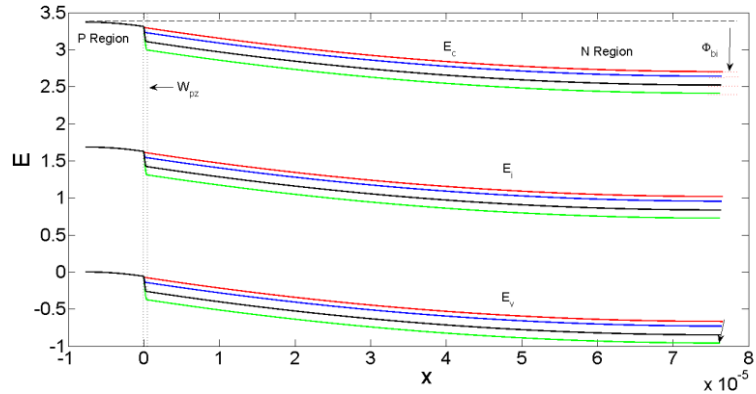

Figure 3. Energy band of the studied PN junction under different strains. From red to green, strains are 0.002, 0.005, 0.008 and 0.01.

Based on these results, in the followed section, we will consider the cases when the PN diode is forward-biased and applied a varied compressive stress at the same time. Admittance is the most important index in dynamic analysis of the PN junction. Therefore, we will study the small-signal admittance brought by the varying piezoelectrical charges that are generated by the time-varying compressive stress *F*.

### 3.1 Low Frequency---small single model

First, under the condition of thermal-equilibrium the relationship between minority carriers concentration on the n- part of the junction and majority carriers concentration on the p-side is given by:

$$p_{n0} = p_{p0}\exp(\frac{-e\varphi_{bi0}}{kT}) \qquad (11)$$

where, $\varphi_{bi0}$ is original built-in potential without applying any force and voltage, $p_{n0}$ is hole concentration in n-side of the junction, $p_{p0}$ is the hole concentration in p-side, $k$ is Boltzmann constant and T is thermal temperature. When the $V_{dc}$ is applied to the junction, the thermal-equilibrium is broken and the hole concentrations on n-side becomes:

$$p_n = p_{p0}\exp(\frac{-e(\varphi_{bi0}-V_{dc})}{kT}) \qquad (12)$$

Now consider the case when there is also a time varied compressive strains existing, we can write the $p_n$, as:

$$p_n = p_{p0}\exp(\frac{-e(c_1+c_2-c_3\sin\omega t+c_4\sin^2\omega t-V_{dc})}{kT}) \qquad (13)$$

where we have taken equation (8) into consideration. The equation (13) can be further simplified as:

$$p_n(x=0) = p_{p0}\exp\left(\frac{e(V_{dc}'+v_1(t)-v_2(t))}{kT}\right) = p_n(0,t) \qquad (14)$$

where $V_{dc}'$=$V_{dc}$-$c_1$-$c_2$, $v_1(t)$=$c_3\sin\omega t$ and $v_2(t)$=$c_4\sin^2\omega t$. If we assume $V_{dc}'$=$V_{dc}''$-$\varphi_{bi0}$, the equaiton (14) can be rewritten as:

$$p_n(0,t) = p_{dc}''\exp(\frac{e(v_1(t)-v_2(t))}{kT}) \approx p_{dc}''(1+\frac{v_1(t)-v_2(t)}{V_t}) \qquad (15)$$

where $p_{dc}''$=$p_{n0}\exp(eV_{dc}''/kT)$, $kT/e$=$V_t$, equaiton (11) and Taylor expansion of the exponential term have been applied. $|v_1(t)- v_2(t)|$<<$V_t$ have been assumed. Equation (15) will be taken as the boundary condition. Due to $c_4$<<$c_3$ in PN junction, the $v_2(t)$ will be ignored in following analysis.

The electric field in the neutral *n* region is assumed to be zero. Thus, the behavior of the excess minority carrier holes that flows from p region can be described by:

$$D_p\frac{\partial^2(\delta p_n)}{\partial x} - \frac{\delta p_n}{\tau_{p0}} = \frac{\partial \delta p_n}{\partial t} \qquad (16)$$

Since the AC voltage that actually comes from the time-varying component of the built-in potential can be seen to superimpose on the DC level, we can write the $\delta p_n$=$\delta p_0(x)$+$p_1 e^{j\omega t}$, where $p_1$ is magnitude of the AC component of the excess concentration. Substituting $\delta p_n$ into equation (16), according to a standard small signal analysis for PN junction, the DC component of the holes diffusion current density can be given by:

$$J_{p0} = \frac{eD_p p_{n0}}{L_p}\left[\exp\left(\frac{eV_{dc}''}{kT}\right) - 1\right] \qquad (17)$$

where $L_p^2$=$D_p\tau_{p0}$ and $C_p^2$=$(1+j\omega\tau_{p0})/ L_p^2$, $\tau_{p0}$ is the hole carriers lifetime. The current density phasor for the sinusoidal component of the diffusion current density is given by:

$$\hat{J}_p = eD_p C_p \left[p_{dc}''\left(\frac{\hat{V}_1}{V_t}\right)\right] e^{-C_p x}\bigg|_{x=0} \qquad (18)$$

The equation (18) can be further written as:

$$\hat{J}_p = J_{p0}\sqrt{1+j\omega\tau_{p0}}\left(\frac{\hat{V}_1}{V_t}\right) \qquad (19)$$

where $J_{p0}=eD_p p_{dc}''$. Likewise, we can derive the current density phasor for minority carrier electrons in p region, which is given by:

$$\hat{J}_n = J_{n0}\sqrt{1+j\omega\tau_{p0}}\left(\frac{\hat{V}_1}{V_t}\right) \tag{20}$$

Combining equations (19) and (20), the PN junction admittance can be derived. We obtain:

$$Y = \frac{A\hat{J}_p + A\hat{J}_n}{\hat{V}_1} = \left(\frac{1}{V_t}\right)\left[I_{p0}\sqrt{1+j\omega\tau_{p0}} + I_{n0}\sqrt{1+j\omega\tau_{n0}}\right] \tag{21}$$

If the AC signal induced by dynamic compressive stress is not large, i.e. $\omega\tau_{p0} \ll 1$ and $\omega\tau_{n0} \ll 1$ (indicating $\omega$ is as $\tau$ approximates to $10^{-7}$ s). The Eq. (21) can be written as: $Y = g_d + j\omega C_d$, where $g_d$ and $C_d$ are diffusion conductance and diffusion capacitance, respectively. Specifically, $g_d = (1/V_t)(I_{p0}+I_{n0})$ and $C_d = (1/2V_t)(I_{p0}\tau_{p0}+I_{n0}\tau_{n0})$.

Followed on these analysis, we have plotted the equivalent circuit diagram of the studied PN junction when there is dynamic compressive forces applied. Generally, the $C_d$ is replaced by a series of piezoelectric diffusion capacitance, i.e. $C_{pd0}$, $C_{pd1}$, $C_{pd2}$,… It is true that all the periodic dynamic external forces applied can be expressed via Fourier expansion with each sinusoidal term leading to one-to-one corresponding diffusion capacitance. $C_j$ is junction capacitance. $r_d$ is resistor of the PN junction and $r_s$ represents the resistor of the external circuit. $I_D$ is the external current. Based on the above analysis, the admittance $Y$ under dynamic force with the frequency varying in [1 Hz, 1 KHz] is numerically calculated, where $g_d$ and $C_d$ are approximately to be 16.8 Ω and 2.97 nF, respectively. Under small signal model, the dynamic compressive force introduces DC components $c_1$ and $c_2$ which modulate the external voltage $V_{dc}$, and therefore changing the current $I_{n0}$ and $I_{p0}$ that are closely related to $Y$.

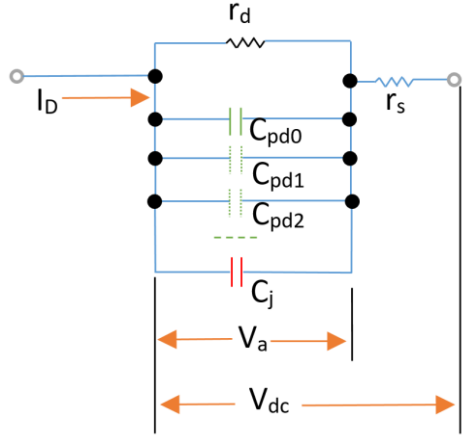

Figure 4. General equivalent circuit diagram of the studied PN Junction under small signal assumption

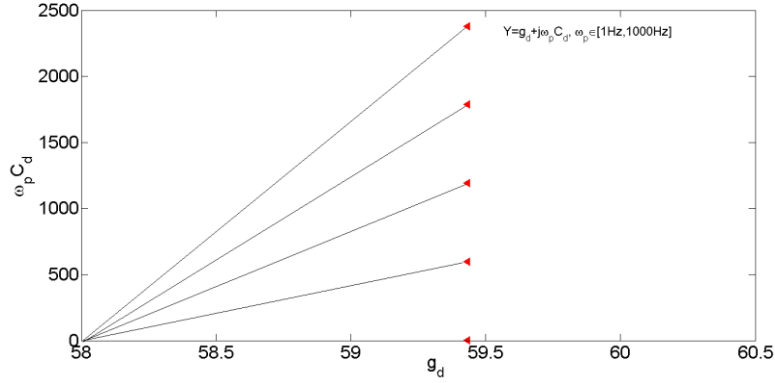

Figure 5. Admittance calculation when external compressed force with frequency varying in [1 Hz, 1 KHz]

**3.2 High Frequency**

For this case, we employed a unified approach [21] for analyzing the dynamic characteristics caused by a dynamic compressive force. When the compressive force is applied at a high frequency, the excess carriers concentration in the n region is then expected to be a more general form:

$$\delta p_n(x,t) = \sum_{k=-\infty}^{\infty} \delta p_{nk}(x) e^{ik\omega t} \quad (22)$$

Inserting above expression into equation (16) taking into account the orthogonality of harmonics, the general solutions for δp$_{nk}$(z) and δn$_{pk}$(z) can be derived:

$$\delta p_{nk}(x) = C_{pk}^{+} \exp\left(-\Lambda_{pk}\frac{x-W_n}{L_p}\right) + C_{pk}^{-}\exp\left(-\Lambda_{pk}\frac{x-W_n}{L_p}\right) \quad (23)$$

where $C_{pk}^{+}$ and $C_{pk}^{-}$ are constants for each harmonics and $\Lambda_{pk}=a_{pk}+ib_{pk}$ with:

$$a_{pk} = \frac{1}{\sqrt{2}}\sqrt{1+\sqrt{1+(k\omega\tau_p)^2}}, \quad b_{pk} = \frac{k\omega\tau_p}{2a_{pk}} \quad (24)$$

When the thickness of the neutral part of the PN junction is much larger than the width of the depletion, i.e. $d_n \gg W_n$, the $C_{pk}^{-}$ in equation (23) should be 0 so that Re$\Lambda_{pk}=a_{pk}$ can be satisfied. The constant $C_{pk}^{+}$ in equation (23) can be found by applying conventional injection boundary conditions $\delta p_{pn}(W_n, t)=p_{n0}f(t)$, in which $f(t)=\exp(qv(t)/kT)-1$ with $v(t)=p_{dc}^{''}+v_1(t)-v_2(t)$. Then $C_{pk}^{+}$ is given by:

$$C_{pk}^{+} = p_{n0}F_k, \quad F_k = \frac{1}{2\pi}\int_{-\pi}^{\pi}f(t)e^{-ik\omega t}d\omega t \quad (25)$$

Substituting $f(t)$ into $F_k$, it arrives:

$$F_0 = I_0(\beta V_\sim)\exp(\beta V_0) - 1; F_k = F_{-k} = I_k(\beta V_\sim)\exp(\beta V_0) \quad (26)$$

where in equation (26) the modified Bessel functions have been introduced, which is defined by:

$$I_k(\beta V_\sim) = \frac{1}{\pi}\int_0^{\pi}e^{\beta V_\sim \cos\omega t}\cos k\omega t\, d\omega t. \quad (27)$$

In equations (26) and (27), β=e/kT, $V_\sim$ represents the amplitude of the time-varying signal that is created by time varied compressive stress. Taking equations (23-27) together into consideration, the δp$_n$(x,t) is finally given by:

$$\delta p_n(x,t) = p_{n0}\sum_{-\infty}^{\infty}F_k\exp(-\Lambda_{pk}\frac{x-W_n}{L_p})e^{ik\omega t} \quad (28)$$

Likewise, the excess carrier of electron in p region also can be derived, which is given by:

$$\delta n_p(x,t) = n_{p0}\sum_{-\infty}^{\infty}F_k\exp(\Lambda_{nk}\frac{x+W_n}{L_n})e^{ik\omega t} \quad (29)$$

The current flowing through the junction is then given by:

$$J(t) = -qD_p \frac{\partial \delta p_n}{\partial x}\bigg|_{x=W_n} A + qD_n \frac{\partial \delta n_p}{\partial x}\bigg|_{x=-W_p} A \qquad (30)$$

Based on the work [21] the dynamic characteristics of the PN junction, i.e. dynamic conductance $g_d$ and diffusion capacitance $C_d$, and for the P$^+$N type junction we have:

$$g_d(\omega, V_\sim) = \frac{\sqrt{1+\sqrt{1+\omega^2\tau_p^2}}}{\sqrt{2}} \frac{I_1(\beta V_\sim)}{\beta V_{\sim/2}} g_{d0}(V_0) \qquad (31)$$

$$C_d(\omega, V_\sim) = \frac{\sqrt{2}}{\sqrt{1+\sqrt{1+\omega^2\tau_p^2}}} \frac{I_1(\beta V_\sim)}{\beta V_{\sim/2}} C_{d0}(V_0) \qquad (32)$$

where $g_{d0}(V_0)=2/\tau_p C_{d0}(V_0)=J_s \exp(\beta V_0)/(kT/e)$. In this work, the $V_0=V_{dc}''$ and $V_\sim \approx c_3$, with $c_4$ assumed to be a much smaller value in our case. By setting k=0, we have DC current $J_0$ from equation (30). It is noted that the dynamic compressive force impacts the $J_0$ as well as dynamic part of overall external current.

Numerically, we have investigated the DC current $J_0$ under compressive force with different amplitudes, indicated by different initial piezoelectric charge width $W_{pz1}$, in Figure 7, to reveal the static current contribution coming from dynamic compressive force applied. As shown in Figure 7, $V_{dc}$ is taken in the range of [0, 1.3V], however, due to the dynamic built-in potential adding in other DC voltage terms, i.e. $c_1$ and $c_2$, the actual voltage applied is $V_{dc}''$, which is changing with different compressive forces. Specifically, as the initial piezoelectric charge width corresponding to the amplitude of $F_1$ increased, the $J_0$ is increased with the threshold for on starting in advance. Meanwhile, the range of $V_{dc}''$ is shifted towards the negative part of the axis.

Subsequently, we have calculated the dynamical characteristics, i.e. diffusion captaincy $C_d$ and diffusion conductance $G_d$, in Figure 8. As the the $C_d$ and $G_d$ are closely related to the frequency ($\omega_p$) of dynamic compressive force, we then vary the $\omega_p$ in high range [1x10$^7$, 1x10$^{10}$] to reveal the frequency dependence of $C_d$ and $G_d$ in Figures (8a) and (8b) respectively. Under different dynamic compressive forces that bring different initial stains $\varepsilon_0$, it is obvious that $G_d$ is increased with $\omega_p$ and sensitive to the initial dynamic compressive strains. The $C_d$, on the other hand, shows a different trend, sensitive to initial dynamic compressive strains though, decreasing sharply at the beginning range of $\omega_p$ and tending to be saturated as $\omega_p$ going higher.

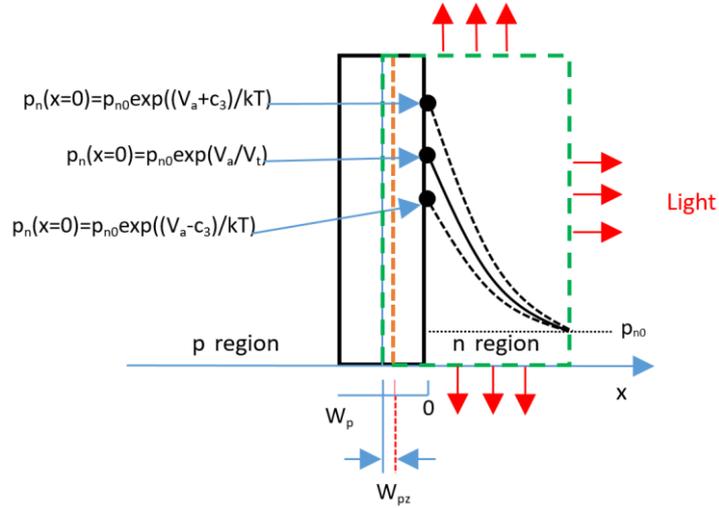

Figure 6. Minority carriers (holes) concentration changes after diffuses into n region with dynamical compressed force applied. Red box indicates the active region for light emitting in piezophotonics.

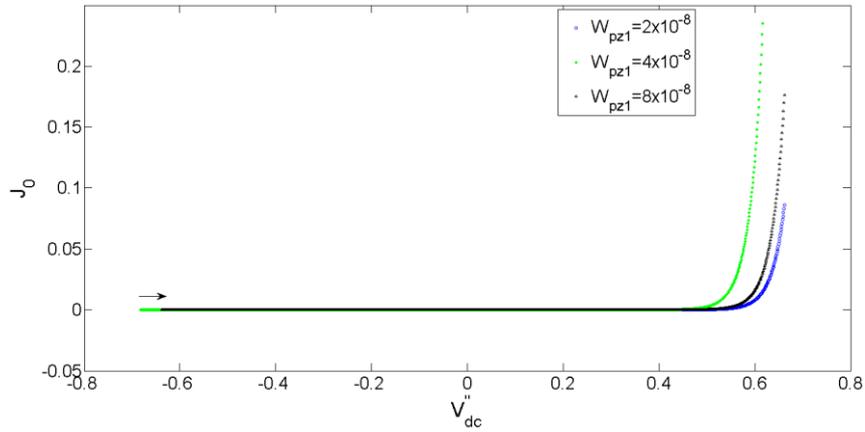

Figure 7. Static current–voltage calculation of the PN junction when dynamic compressive force applied. Three different amplitude of compressive force are discussed.

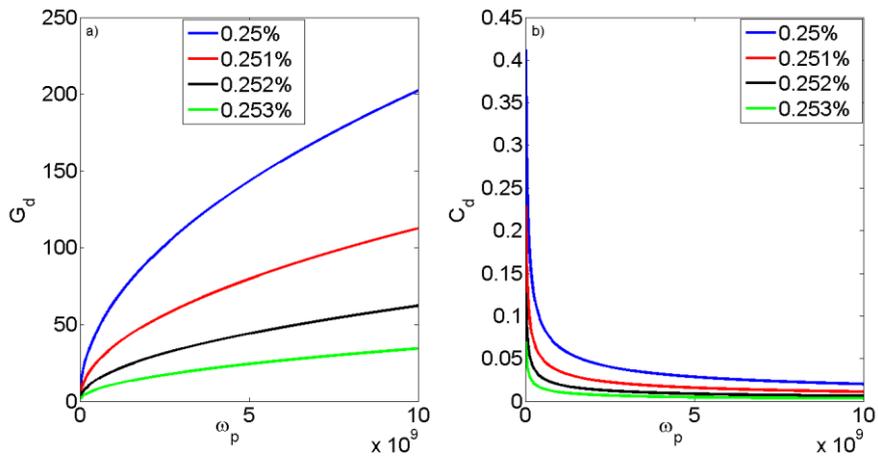

Figure 8. Dynamical diffusion capacitance and conductance subjected to different dynamic compressive strains are calculated.

## 4. Dynamical analysis of recombination process in piezotronics and piezo-phototronics

The piezotronic effect affects the recombination process happening in the depletion region, and the recombination process subsequently affects the carrier's transportation and electron-photon interactions in piezotronics and piezo-phototronics. Therefore, understanding the mechanism of the recombination subject to the piezotronic effect is essential for both piezoelectronics and piezophotonics. In the PN junction we studied in previous sections with one of its n-part made by the ZnO, the recombination process happening in the depletion region is mainly attributed to the defect assisted bulk and surface recombination-generation process, which is given by:

$$R = \frac{C_n C_p N_t (np - n_i^2)}{C_n(n+n') + C_p(p+p')} \quad (33)$$

By taking $n=n_i exp((E_{fn}-E_{fi})/kT)$ and $p=n_i exp((E_{fi}-E_{fp})/kT)$, the recombination rate $R$ in depletion region suffering a different amplitude of compressive force can be calculated, to reveal how the piezotronic effect affects the recombination process. The numerical results are shown in Figure 9. It is seen that the $R$ has a peak around the interface of PN junction, and the peak increases accordingly as the compressive strain increases. The position of the maximum $R$ is shifted slightly to the right direction when the compressive strain is increased. Dynamically, we have also calculated maximum $R$ subject to dynamic compressive force with different frequencies $f$ (f=1 Hz and f=10 Hz) in Figure 10. When there is a dynamic compressive force with certain frequency is applied, the recombination rate $R$ is dynamically changed with the same pattern. The result gives a quantitative analysis to the impacts brought by piezotronic effect to the recombination and transportation process.

In a typical PN junction, the total forward-bias current is the sum of recombination and the diffusion current. Therefore, different external compressive stresses will lead to the different recombination rates, and in turn modulating the total current as well as the light emitting intensity in piezotronic LED. For illustrating on how the dynamic compressive force modulates the light emitting intensity in piezoelectronic LEDs, we have conducted the following analysis.

As shown in Figure 6, we treat the dotted box as active region in which light emitting and recombination are happening and related. In this region, the rate equation of carrier hole can be modeled as[22] [23]:

$$\frac{dp}{dt} = \frac{\eta_i I}{eV_a} - \left[A(p-p_i) + B(p^2 - p_i^2) + Cp(p^2 - p_i^2)\right] \quad (34)$$

where $A$ is coefficient of defect assisted recombination. $B$ and $C$ are the coefficients of radiative recombination-generation and Auger recombination, respectively. $p_i$ is intrinsic electron concentration. $\eta_i$ is the fraction of the total current that is due to the recombination in the active region. $V_a$ is the volume of the active region. $I$ is the total current flowing through the external circuit in forward bias. Taking the parameters as: A=1x10$^7$, B=1x10$^{-9}$, C=5x10$^{-29}$, $N_A$=1x10$^{17}$/cm$^3$, $N_D$=1x10$^{15}$/cm$^3$, initial piezoelectric stain $\varepsilon_0$=0.02/100, and by fixing I=0.1 A, we have calculated the number of photons emitted per second in $V_a$. The result is shown in Figure 11, where six different frequencies of dynamic compressive force are studied. The light emitting in piezophotonics is closely related to the dynamic force applied.

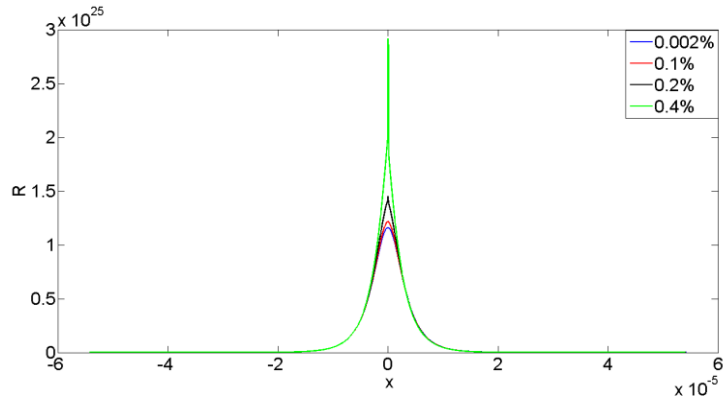

Figure 9. Recombination rated R changes with different stains

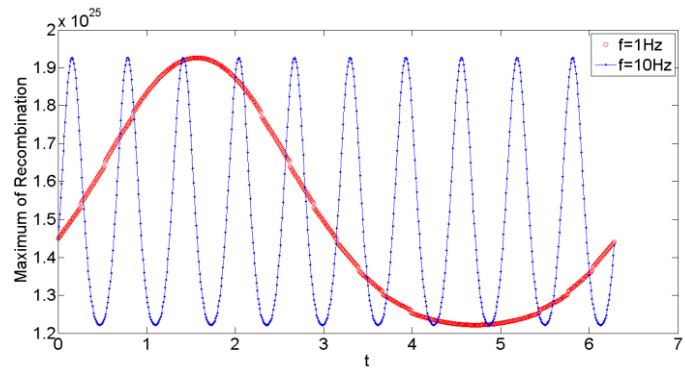

Figure 10. Dynamic recombination rate when there is dynamic compressive force with frequency 1 Hz and 10 Hz applied.

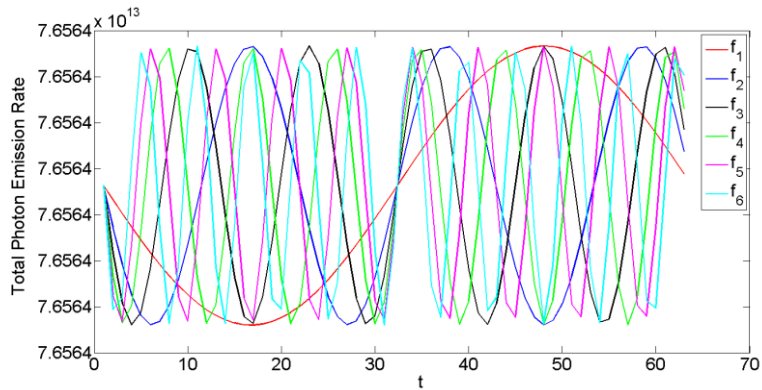

Figure 11. Photons emitted per second in active region when dynamic compressive force with different frequencies applied.

**Conclusion**

To conclude, dynamic built-in potential exists in the piezotronic PN junction when a dynamic compressive stress is applied. The dynamic built-in potential contributes both DC and AC parts together with the applied external voltage, which can modulate current-voltage characteristics, threshold, diffusion capacitance and conductance. The work has thoroughly investigated these dynamic characteristics under external compressive stresses at both low and high frequencies by developing a small signal model and united approach, respectively. The results are self-consistent and provide a general procedure for treating dynamic regime of piezotronics. The piezotronic PN junction is the fundamental element of the piezo-phototronic LEDs. The work has quantitatively calculated the dynamic recombination rate and light emitting intensity for the device. Unlike the previous work, this work presents detailed results assisting in understanding how the piezo-potential affects the dynamic process of carriers in the piezo-phototronic LEDs.

**References**

[1]     Z. L. Wang and W. Z. Wu, "Piezotronics and piezo-phototronics: fundamentals and applications," *National Science Review,* vol. 1, pp. 62-90, Mar 2014.

[2]     Z. L. Wang and J. H. Song, "Piezoelectric nanogenerators based on zinc oxide nanowire arrays," *Science,* vol. 312, pp. 242-246, Apr 14 2006.

[3]     W. Z. Wu*, et al.*, "Piezoelectricity of single-atomic-layer MoS2 for energy conversion and piezotronics," *Nature,* vol. 514, pp. 470-+, Oct 23 2014.

[4]     Z. L. Wang, "Piezopotential gated nanowire devices: Piezotronics and piezo-phototronics," *Nano Today,* vol. 5, pp. 540-552, 2010.

[5]     M. Z. Peng*, et al.*, "High-Resolution Dynamic Pressure Sensor Array Based on Piezo-phototronic Effect Tuned Photoluminescence Imaging," *Acs Nano,* vol. 9, pp. 3143-3150, Mar 2015.

[6]     R. R. Bao*, et al.*, "Flexible and Controllable Piezo-Phototronic Pressure Mapping Sensor Matrix by ZnO NW/p-Polymer LED Array," *Advanced Functional Materials,* vol. 25, pp. 2884-2891, May 20 2015.

[7]     Y. Liu*, et al.*, "Nanowire Piezo-phototronic Photodetector: Theory and Experimental Design," *Advanced Materials,* vol. 24, pp. 1410-1417, Mar 15 2012.

[8]     C. F. Pan*, et al.*, "Enhanced Cu2S/CdS Coaxial Nanowire Solar Cells by Piezo-Phototronic Effect," *Nano Letters,* vol. 12, pp. 3302-3307, Jun 2012.

[9]     P. Bonato, "Wearable Sensors and Systems From Enabling Technology to Clinical Applications," *Ieee Engineering in Medicine and Biology Magazine,* vol. 29, pp. 25-36, May-Jun 2010.

[10]    W. W. a. Z. L. Wang, "Piezotronics and piezo-phototronics for adaptive electronics and optoelectronics," *NATURE REVIEWS MATERIALS* vol. 1, pp. 1-17, 2016.

[11]    D. H. Kim, "Epidermal electronics (vol 333, pg 838, 2011)," *Science,* vol. 333, pp. 1703-1703, Sep 23 2011.

[12]    Y. Zhang*, et al.*, "Fundamental Theory of Piezotronics," *Advanced Materials,* vol. 23, pp. 3004-3013, Jul 19 2011.

[13]    W. Liu*, et al.*, "First principle simulations of piezotronic transistors," *Nano Energy,* vol. 14, pp. 355-363, May 2015.

[14]    L. S. Jin and L. J. Li, "Quantum simulation of ZnO nanowire piezotronics," *Nano Energy,* vol. 15, pp. 776-781, Jul 2015.

[15]    W. Liu*, et al.*, "Density functional studies on edge-contacted single-layer MoS2 piezotronic transistors," *Applied Physics Letters,* vol. 107, Aug 24 2015.

[16]    A. Rinaldi*, et al.*, "The Clash of Mechanical and Electrical Size-Effects in ZnO Nanowires and a Double Power Law Approach to Elastic Strain Engineering of Piezoelectric and Piezotronic Devices," *Advanced Materials,* vol. 26, pp. 5976-+, Sep 10 2014.

[17]    M. X. Chen*, et al.*, "Tuning Light Emission of a Pressure-Sensitive Silicon/ZnO Nanowires Heterostructure Matrix through Piezo-phototronic Effects," *Acs Nano,* vol. 10, pp. 6074-6079,

Jun 2016.

[18] Z. N. Wang, *et al.*, "Ultrafast Response p-Si/n-ZnO Heterojunction Ultraviolet Detector Based on Pyro-Phototronic Effect," *Advanced Materials,* vol. 28, pp. 6880-+, Aug 24 2016.

[19] S. Kim, *et al.*, "Piezoelectric effects on carrier capturing time in a hybrid p-n junction system: a numerical study using finite element method," *Journal of Computational Electronics,* vol. 15, pp. 40-44, Mar 2016.

[20] X. Y. Li, *et al.*, "Enhancing Light Emission of ZnO-Nanofilm/Si-Micropillar Heterostructure Arrays by Piezo-Phototronic Effect," *Advanced Materials,* vol. 27, pp. 4447-4453, Aug 12 2015.

[21] A. A. Barybin and E. J. P. Santos, "A unified approach to the large-signal and high-frequency theory of p-n-junctions," *Semiconductor Science and Technology,* vol. 22, pp. 1225-1231, Nov 2007.

[22] J. Singh, *Electronic and Optoelectronic Properties of Semiconductor Structures* vol. 34: Springer Netherlands, 2008.

[23] N. P. SL Chuang, S Koch,, *Physics of Optoelectronic Devices*: Wiley, 2008.

**Acknowledgements**

National Natural Science Foundation of China (61604078), Natural Science Youth Foundation of Jiangsu Province (BK20160905). Natural Science Foundation of the Jiangsu Higher Education Institutions of China (16KJB510029), Postdoctoral Science Foundation of China (2017M610341), Jiangsu Natural Science Foundation for Excellent Young Scholar (BK20170101) and NUPTSF (NY216010) are acknowledged. SPARC II (Solar Photovoltaic Academic Research Consortium) project funded by the Welsh European Funding Office is acknowledged.

**Additional information**

Competing financial interests: The authors declare no competing financial interests.